\documentclass[epj]{svjour}

\usepackage{graphics}
\usepackage{epsfig}

\begin{document}

\title{Non-linear Osmotic Brush Regime: Simulations and mean-field theory 
\thanks{A revised version of Ref. \cite{Naji03} with corrections
in the simulation part from C. Seidel, Macromolecules {\bf 38}, 2540 (2005). }
       }

\author{Ali Naji
        \inst{1,2,}\thanks{\email{naji@ph.tum.de}}
        \and 
        Roland R. Netz
        \inst{1,2,}\thanks{\email{netz@ph.tum.de}}
        \and
        Christian Seidel
        \inst{2,}\thanks{\email{seidel@mpikg-golm.mpg.de}}
}

\institute{Physics Department, Technical University of Munich, 
        D-85748 Garching, Germany. 
           \and 
         Max-Planck Institute of Colloids and Interfaces, 
         Am M\"uhlenberg, D-14476 Golm, Germany.
           }

\date{Received: date / Revised version: date}

\abstract{
We investigate polyelectrolyte brushes in the osmotic regime 
using both theoretical analysis and molecular dynamics simulation
techniques. 
In the simulations at moderate Bjerrum length, 
we observe that the  brush height varies weakly with 
grafting density, in contrast to the accepted 
scaling law, which predicts a brush thickness independent of 
the grafting density. We show that such behavior can be explained
by considering lateral electrostatic effects (within 
the non-linear Poisson-Boltzmann theory) combined with 
the coupling between lateral and  
longitudinal degrees of freedom due to the
conserved polymer volume (which are neglected in scaling arguments). 
We also take the non-linear elasticity 
of polyelectrolyte chains into consideration, which makes significant effects 
as chains are almost fully stretched in the osmotic regime. It is shown that  all these
factors lead to a non-monotonic behavior
for the brush height as a function of the grafting density. 
At large grafting densities, the brush height increases with increasing the  
grafting density due to the volume constraint. 
At small grafting densities, we obtain a 
re-stretching of the chains for decreasing 
grafting density, which is caused by lateral electrostatic contributions
and the counterion-condensation process
at polyelectrolyte chains. These results are obtained assuming all counterions
to be trapped within the brush, which is valid for sufficiently long chains of 
large charge fraction. 
}

\maketitle




\section{Introduction}
\label{sec:intro}

In recent years, polyelectrolyte brushes have been subject of 
extensive  investigations both theoretically \cite{MIK88,MIS89,PIN91,BOR91,ROS92,ZHU92,WIT93,ISR94,BOR94,PRY96,ZHU97,AMOS95,CSA01,CSA99,CSA00,CS02} 
and experimentally \cite{MIR95,GUE95,AHR97,AHR98,MUL00,MUL01,GUE98,BAL02,ABE02,HELM,GUENOUN}. 
Polyelectrolyte brushes are layers of charged polymer chains 
densely end-grafted onto surfaces of various geometries. 
They have important technological applications notably in 
stabilization of colloidal dispersions in polar media, where
forces between colloidal particles (coated by grafted 
polyelectrolytes) are modified and may be controlled in order 
to prevent flocculation \cite{NAP83,PAR97}.
Recent investigations \cite{MIK88,MIS89,PIN91,BOR91,ROS92,ZHU92,WIT93,ISR94,BOR94,PRY96,ZHU97,AMOS95,CSA01,CSA99,CSA00,CS02,MIR95,GUE95,AHR97,AHR98,MUL00,MUL01,GUE98,BAL02,ABE02,HELM,GUENOUN}
have revealed the detailed phase behavior of 
charged brushes over a wide range of
system parameters, namely, grafting density and charge 
fraction of chains. (Here we 
consider only planar brushes with no added salt.)  
It is known from scaling theories as well as from
self-consistent field studies that the interplay between steric, 
entropic and electrostatic contributions results in a variety 
of different scaling relations for the equilibrium height of the brush.

In strongly charged brushes with a large effective charge density 
of grafted polyelectrolytes,
{\em i.e.} when both grafting density and charge fraction of
polyelectrolyte chains are sufficiently large, effectively all 
of the counterions 
are trapped inside the 
brush. Over a certain range of  charge fractions and grafting densities, 
the repulsive osmotic 
pressure of counterions is the major effect that tends to swell 
the chains and balances their elastic stretching pressure. This regime 
is known as the osmotic brush regime, where the
equilibrium thickness of the brush was obtained by scaling arguments
to be independent of the grafting density \cite{PIN91,BOR91}.  
The osmotic brush height may, however, exhibit a weak 
dependence on the grafting density
as a result of non-uniform counterion  distribution  inside the brush. 
Specifically, variation of the counterion density profile in the direction
normal to the anchoring
plane (which may be caused by partial diffusion
of counterions outside the brush layer) leads to a 
logarithmic dependence on the grafting density \cite{ZHU97}.
A weak dependence of the brush height on grafting density 
may also be produced by lateral inhomogeneities 
as will be addressed in this paper.  The crossover between these 
two mechanisms is regulated by the chain charge fraction and 
grafting density. 
In the regime of large grafting densities, 
steric effects dominate over electrostatic interactions and the 
excluded-volume repulsion (in good solvent conditions) balances the
elastic pressure, leading to the so-called quasi-neutral brush 
regime \cite{BOR91,BOR94}.  The effects of excluded-volume interactions 
have also been studied in poor solvent conditions \cite{ROS92}. 
On the other hand,  for larger charge fractions, the electrostatic 
correlations due to Debye-H\"uckel interactions between mobile 
counterions and oppositely charged monomers may produce a 
dominant attractive pressure balancing the excluded-volume 
repulsion \cite{BOR91,CSA01}. 
This leads to a collapsed brush regime, which has 
also been obtained in previous molecular dynamics simulations 
\cite{CSA01,CSA99,CSA00}.

In weakly charged brushes (which are not considered in this paper),
the counterion cloud extends far beyond the brush height and the 
osmotic pressure of the counterions becomes irrelevant against
uncompensated electrostatic repulsion between charged monomers. 
 This results in  the charged or Pincus brush regime, where the elastic 
and the electrostatic pressures are balanced \cite{PIN91}. Details of
 the phase diagram and transitions between these regimes have been 
discussed extensively in the literature 
\cite{MIK88,MIS89,PIN91,BOR91,ROS92,ZHU92,WIT93,ISR94,BOR94,PRY96,ZHU97,AMOS95,CSA01}. 

In recent molecular dynamics simulations \cite{CS02}, 
structure and equilibrium properties of salt-free
planar osmotic brushes have been studied at moderate Bjerrum length.
For moderate values of charge fraction and 
grafting density, it has been observed that the 
counterions are mostly confined within the brush and generate an
almost step-like density profile in the direction normal to the anchoring 
plane. 
In contrast, counterions exhibit a non-uniform distribution in lateral
directions. In this situation, the simulated brush 
thickness varies weakly with the grafting density in contrast to  
the standard scaling law  \cite{PIN91,BOR91}.

In scaling studies, it is commonly assumed that the counterions 
are distributed uniformly in lateral directions parallel to the 
anchoring plane and  that the elasticity of chains is obtained from 
a linear (Gaussian) model. 
In this paper, we present a mean-field
model for the osmotic brush, in which 
we consider corrections to both of these assumptions:
Firstly, we take into account {\em lateral} electrostatic effects including  
lateral variation of the counterion density profile around polyelectrolyte
chains within the framework of 
the non-linear Poisson-Boltzmann (PB) theory, and secondly, 
we use a freely-jointed-chain model to mimic non-linear elastic stretching of
the polyelectrolyte chains.
Moreover, we take into account 
excluded-volume interactions between particles as well as the
conserved polymer volume as the chain dimensions change, as will be
described below. 
The results of our model for the behavior of the brush 
thickness, which we refer to as the
{\em non-linear osmotic brush regime}, display
a non-monotonic dependence on the grafting density. In  particular, 
at moderate grafting densities,
we find that the brush height weakly increases
with the grafting density arising as a result of the interplay between
lateral electrostatics and  the coupling effects. This behavior will be 
compared with simulation results
displaying a reasonable agreement.
A weak dependence of the brush height on grafting
density has also been observed in recent experiments  \cite{HELM,GUENOUN} 
and has been compared with the non-linear osmotic brush predictions 
obtained using simple scaling arguments  \cite{HELM}. 
At small grafting densities, our model predicts re-stretching of the
chains by lowering the grafting density, which is caused by 
an increasing electrostatic pressure acting on the chains in this limit. 
As we shall demonstrate, this behavior is regulated by the
counterion-condensation process
around the chains, which is captured within the non-linear PB theory. 
This regime has not been investigated in simulations
and experiments 
yet, and we propose that one needs long chains to obtain this behavior
(see the Discussion).

An inhomogeneous distribution of counterions in lateral directions 
has been recently 
observed both in simulation \cite{CSA99,CSA00,CS02} and experiment 
\cite{AHR98,MUL01,BAL02,ABE02}. Analyzing experimental data on spherical charged brushes, 
Muller {\em et al.} \cite{MUL01}  observe that the counterion distribution 
around polyelectrolyte chains indeed follows the 
the non-linear PB predictions as obtained within the cylindrical-cell-model approach 
\cite{ALFREY,FUOSS}. 
Other experiments indicate that a large fraction of counterions 
binds strongly to the polyelectrolyte chains, thus counterions are not
evenly distributed inside the brush \cite{AHR98,BAL02,ABE02}.
On the other hand, the
non-linear elasticity, which accounts for the finite 
extensibility of the chains, appears to be essential for
strongly charged brushes at moderate grafting densities, as 
polyelectrolyte chains in these situations are found to be stretched up to 
60-80$\%$ of their contour length \cite{CS02,GUE95,AHR97,AHR98,MUL01,BAL02,HELM,GUENOUN}. 
Non-linear elasticity models have also been used in previous works 
\cite{PRY96} (but without considering lateral effects
as done here).
It is important to note that the non-linear elasticity by itself 
can not generate a grafting-density dependence for the brush height
\cite{AMOS95}.

We shall neglect the variation of the counterionic 
density profile in the direction normal to the anchoring plane 
and assume that all counterions are trapped within the brush. 
In the range of parameters, 
where a sizable fraction of counterions could leave the brush layer, the step 
form of the density profile in the normal direction is changed
and may be calculated using self-consistent field (SCF) techniques 
\cite{MIK88,MIS89,ZHU92,ISR94,PRY96,ZHU97}. 
The SCF study of Zhulina and Borisov \cite{ZHU97} indeed shows that
at large grafting densities (osmotic regime), 
the variation of the density profile in the normal direction gives rise to a logarithmic 
dependence of the brush height on the grafting density. 
Such a dependence is obtained by treating the electrostatic effects
at the non-linear level using the PB equation but assuming that
the concentration of counterions (and that of monomers) is 
smeared in lateral directions 
\cite{ZHU97}. Our model, therefore, describes the complementary limit, which 
can be achieved by taking sufficiently long chains of large charge fraction
(even at low grafting densities) as will be examined later.

In the present study, we account for the volume interaction
between counterions and polyelectrolyte chains using a closest-approach
distance, but neglect the excluded-volume 
interactions between counterions themselves, which may 
be justified in the considered range of parameters.
In addition, the effective volume of the polyelectrolyte chain,
which is not accessible for counterions, is assumed to be constant:
when chains shrink (at a fixed grafting density), the available
volume for counterions decreases leading to an increasing osmotic 
pressure. 
Such a coupling between longitudinal and 
lateral degrees of freedom is a very simple way to 
mimic the back influence of the conformational changes of the 
chains on the true
volume available for counterions inside the brush and their osmotic pressure.




\section{The geometry of the model}
\label{sec:model}

\begin{figure}[t]
\includegraphics[angle=0,width=8.5cm]{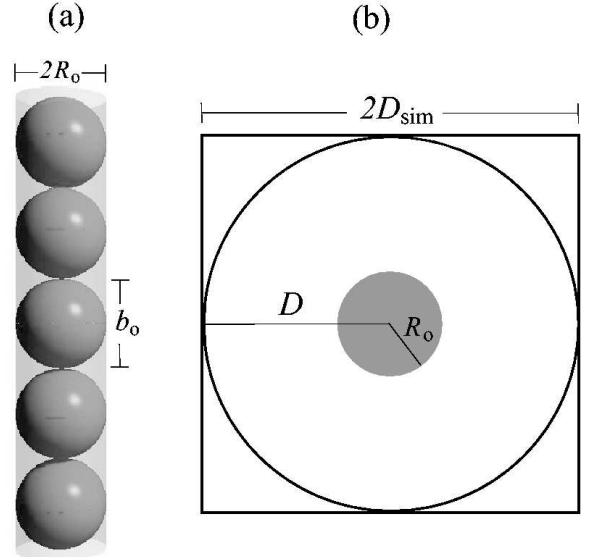}
\smallskip
\caption{\label{fig:cells} The model used for the polyelectrolyte chain (a) and the cylindrical unit 
cell (b) for the brush as discussed in the text. The unit cell boundary in the simulations is shown by a 
square.}
\end{figure}

The model which shall be employed in the present study to 
calculate the electrostatic free energy of the brush is based on 
the commonly-used cell model for rod-like polyelectrolyte solutions 
\cite{KATCH,HOL01}. Each polyelectrolyte chain in the brush 
is assumed to have $N$ spherical monomers of the same 
{\em diameter} $b_0$ among which a fraction of $f$ 
is charged with a charge valency of $q$. 
We assume that each single chain is symmetrically enclosed 
in a cylindrical unit cell with radius $D$, which is determined by the 
grafting density $\rho_{\mathrm{a}}$.  
The chains length, $L$, is assumed to be much larger 
than the radius of the cell, $L\gg D$. 
Furthermore, each polyelectrolyte chain is 
modeled as a cylindrical rod (which we may refer to as 
the polyelectrolyte rod) with the same length and some radius $R \leq D$.  
We suppose that the total volume of the rod is
constant, 
{\em i.e.}, $R^2L=R_0^2L_0$, where $L_0$ and $R_0$ stand for
the length and radius associated with its fully stretched conformation; 
$L_0$ is taken equal to the contour length of the chain, 
$L_0=Nb_0$, but  $R_0$ can be generally different from the 
radius of the monomers due to different possibilities to choose a 
cylindrical rod model for a polyelectrolyte chain \cite{Naji03}. 
We restrict our discussion to a model in which $R_0=b_0/2$ (Figure \ref{fig:cells}a).

The electric charge of the polyelectrolyte chain, $qfN$,
is assumed to be uniformly 
distributed over the surface of the rod with a linear charge density of 
$\tau$ (all in units of elementary charge $e$). 
There is a fixed number of oppositely charged counterions, $N_c$, 
with charge valency of $q_c$
and radius of $r_c$ confined 
inside each unit cell, such that the electroneutrality condition,
$\tau L=qfN=q_c N_c$, is fulfilled. (The charge valencies $q$ and $q_c$, and
also $\tau$, are defined to be positive.) 
Solvent is treated 
as a continuum background characterized completely by
its dielectric constant $\varepsilon$, and 
no additional salt is present in the system.

In the simulation model,
 grafting points form a square lattice with the lattice
 spacing of $2D_{sim}$ and thus, a grafting density of 
$\rho_{\mathrm{a}}=1/(4D^2_{sim})$. 
A natural space-filling choice of unit cell for this system is a 
square unit cell with the edge size of $2D_{sim}$. 
Since the analytical solution for the counterion distribution is
only available for a cylindrical unit cell, we have to map the square 
simulation box to a cylindrical box. (Note that experimentally,
the grafting-point distribution is highly irregular and thus different
from both simulation and analytical model.) 
There are different 
ways to adopt a cylindrical unit cell for such a planar brush 
as discussed in detail in Ref. \cite{Naji03}. Here 
we choose a cylindrical cell with the diameter, 
$2D$, equal to the lattice 
spacing of the square lattice as shown in Figure \ref{fig:cells}b, where $D=D_{sim}$.




\section{Non-linear osmotic brush regime within mean-field approximation}
\label{sec:OsB}




\subsection{The electrostatic free energy}
\label{sub:elec}

The so-called mean-field or Poisson-Boltzmann theory of a
cylindrical cell model for polyelectrolyte solutions 
has been presented long time ago \cite{ALFREY,FUOSS,LIF}. 
In the mean-field approximation, the electrostatic correlations 
between neighboring cells are entirely neglected 
and the study of the system is therefore 
reduced to a single-cell study. In addition to this, correlations between 
counterions present in the same cell are systematically 
neglected, however, they remain
still correlated with the polyelectrolyte rod. 
The system is then studied in its ``ground state'', where all the 
fluctuations are neglected as well. Then, by virtue of 
the electroneutrality condition, it follows 
that the electric field vanishes over the cell boundary.

A canonical field theory may be written for the system
and the 
Poisson-Boltzmann (PB) theory is subsequently obtained as a saddle-point 
approximation to the corresponding action \cite{NET99b}. 
The canonical PB free energy per unit cell 
is calculated and can be written (in units of $k_B T$) as
\begin{eqnarray}
    {\mathcal F}^{PB}_{N_c}&=&
                    -\frac{1}{4\pi \ell_B q_c^2}
                       \int d{\mathbf r}\,
                          [\frac{1}{2}(\nabla \psi)^2
                           +
                           \frac{2\xi}{R}\delta(r-R)\psi({\mathbf r})]
                       \nonumber\\
                        &-&N_c\ln 
                             [\int d{\mathbf r}\,
                                 \Omega({\mathbf r})
                                   e^{-\psi({\mathbf r})}]
                        +{\mathcal C},
\label{eq:PBfree}
\end{eqnarray}
where $\ell_B=e^2/(4\pi \varepsilon\varepsilon_0k_BT$) is the 
Bjerrum length (associated with a medium with dielectric constant of 
$\varepsilon$), $\xi=\ell_B q_c \tau$ is the Manning parameter, and 
${\mathcal C}$ is a constant, which 
contains contributions due to self-energies, kinetic 
energy of counterions and other numerical constants. The geometry function 
$\Omega({\mathbf r})$ takes into account the presence of hard walls and 
 restricts the positions of mobile counterions 
to the cylindrical region of $R\leq r \leq D$, that is 
$\Omega({\mathbf r})=1$ for  $R\leq r \leq D$ and zero otherwise, 
where $r$ denotes the radial distance from the axis of the cylindrical unit cell. 
As a result of the saddle-point approximation,
$\delta {\mathcal F}^{PB}_{N_c}/\delta \psi=0$,   
the dimensionless potential field $\psi({\mathbf r})$ is obtained to 
fulfill the so-called Poisson-Boltzmann (PB) equation 
\begin{equation}
       \nabla^2 \psi=
       \frac{2\xi}{R}\delta(r-R)
                       -\kappa^2\Omega({\mathbf r})
                            e^{-\psi({\mathbf r})},
\label{eq:PB}
\end{equation}
where 
$\kappa^2=4\pi\ell_Bq_c^2N_c/\int d{\mathbf r} \Omega({\mathbf r})\exp(-\psi)$ 
is an unspecified factor the value of which will be fixed once we specify the 
reference point of the potential \cite{Note3}. 
To calculate the 
electrostatic free energy, one needs to solve Eq. (\ref{eq:PB}) for 
the potential field $\psi$. (Here the physical electrostatic potential
reads $k_BT\psi/q_ce$.) 
Assuming that the polyelectrolyte rod is sufficiently long and 
that the solution for the potential field also has 
cylindrical symmetry, it follows that
\begin{eqnarray}
          \left(r\frac{d\psi}{dr}\right)_{r=R}=
                                               2\xi 
                        \,\,\,\,{\textrm{and}}\,\,\,
          \left(r\frac{d\psi}{dr}\right)_{r=D}=0, 
\label{eq:PBboundary}
\end{eqnarray}
where we have used the global electroneutrality condition 
\begin{equation}
  \xi L=N_c\ell_B q_c^2.
\end{equation}
It is easily observed that the 
only non-trivial contribution to the PB free energy comes from the 
potential field in the interior region $R\leq r \leq D$, where
the solution to Eqs. (\ref{eq:PB}) and (\ref{eq:PBboundary}) is available
due to early works by Alfrey {\em et al.} \cite{ALFREY} and Fuoss {\em et al.} 
\cite{FUOSS}. It was shown that the solution for $\psi(r)$ 
takes different functional forms depending on whether $\xi$ 
lies below or above a threshold value $\xi_c$. 
For a polyelectrolyte rod with {\em fixed radius}, $R$,
the threshold is simply given by 
\begin{equation}
  \xi_c=\frac{\ln (D/R)}{1+\ln (D/R)},
\label{eq:xicFuoss}
\end{equation}
while in 
our model with constant volume constraint, as we shall see later,
$\xi_c$ has to be determined from a transcendental equation.

For $\xi \leq \xi_c$, the solution reads
\begin{equation}
     \psi(r)=
             \ln[
                 \frac{\kappa^2r^2}{2\beta^2}
                    \sinh^2(\beta \ln \frac{r}{D}
                               -\tanh^{-1}\beta)],
\label{eq:Solution}
\end{equation}
where $\beta$ is given by the transcendental equation
\begin{equation}
     \xi=
        \frac{1-\beta^2}
                {1-
                   \beta\coth
                              (-\beta\ln \frac{D}{R})}.
\label{eq:beta}
\end{equation}
Inserting the solution Eq. (\ref{eq:Solution}) into the expression
(\ref{eq:PBfree}), we obtain the electrostatic free energy 
per number of monomers 
(up to an additive constant and in units of $k_BT$) as
\begin{eqnarray}
   \frac{{\mathcal F}^{PB}}{N}
               &=&\frac{qf}{q_c}
                   \{-\frac{1}{\xi}
                      [(1+\beta^2)\ln \frac{D}{R}+
                          \ln \left(\frac{(\xi-1)^2-\beta^2}
                            {1-\beta^2}\right)
                \nonumber\\
                &+&\xi]
                     +\ln [(\xi-1)^2-\beta^2]
                     -\ln (2\pi\ell_B f q_c q N R^2)\}.
\label{eq:betafree}
\end{eqnarray}

For  $\xi \geq \xi_c$, the solution to the PB equation is
\begin{equation}
     \psi(r)=
             \ln[
                 \frac{\kappa^2r^2}{2\beta^2}
                    \sin^2(\beta \ln \frac{r}{D}
                               -\tan^{-1}\beta)],
\label{eq:Solution2}
\end{equation} 
where $\beta$ now satisfies the following equation
\begin{equation}
     \xi=
        \frac{1+\beta^2}
                {1-\beta
                        \cot
                            (-\beta\ln \frac{D}{R})}.
\label{eq:beta2}
\end{equation}
Similarly, we can calculate the electrostatic free energy of the system,
\begin{eqnarray}
    \frac{{\mathcal F}^{PB}}{N}
                &=&\frac{qf}{q_c}\{-\frac{1}{\xi}
                        [(1-\beta^2)\ln \frac{D}{R}+
                          \ln \left(\frac{(\xi-1)^2+\beta^2}
                                        {1+\beta^2}\right)
                \nonumber\\     
                &+&\xi]
                        +\ln [(\xi-1)^2+\beta^2]
                        -\ln (2\pi\ell_B f q_c q N R^2)\}.
\label{eq:betafree1} 
\end{eqnarray}
The free energies (\ref{eq:betafree}) and (\ref{eq:betafree1})
have also been derived  by 
Lifson and Katchalsky \cite{LIF} using a charging process.

We note that here $R$ represents the actual 
radius of the polyelectrolyte rod and thus in our model
is related to the 
Manning parameter of the system $\xi$ through
\begin{equation}
      R(\xi)=R_0\sqrt{\xi/\xi_0}.
\label{eq:oldR}
\end{equation}
This dependence is induced by the volume constraint 
$R^2L=R_0^2L_0$, and 
the fact that the total charge of the rod is conserved, that means
\begin{equation}
   \xi L=\xi_0 L_0,
\label{eq:xiL}
\end{equation}
where 
\begin{equation}
  \xi_0=q_cqf\frac{\ell_B}{b_0}
\label{eq:xi0}
\end{equation}
is the Manning parameter 
associated with the fully stretched conformation of the rod. 
In the above formulation, counterions have been taken as point-like 
particles. To account for the finite size of counterions, 
we may define a closest 
approach distance between the rod and counterions, $R_{ca}$,
to be used instead of $R$ in the preceding equations. 
Assuming that the counterions have radius of $r_c$ and recalling
the actual radius of the rod $R$ from Eq. (\ref{eq:oldR}), we have 
\begin{equation}
      R_{ca}(\xi)=
             r_c+R_0\sqrt{\xi/\xi_0}.
\label{eq:R}
\end{equation}
Note that $\xi$ is limited from above and below due to
the geometrical constraints  $ R_{ca}\leq D$ and $L\leq L_0$ respectively. 
As a result,
one observes that 
\begin{equation}
  \xi_0\leq \xi \leq \xi_u,
\label{eq:limitonxi}
\end{equation}
 where
\begin{equation}
        \xi_u=
              \xi_0 
                (D-r_c)^2/R_0^2. 
\label{eq:upxi}
\end{equation}

Finally, we remark that here the threshold Manning parameter, $\xi_c$, at 
which the functional form of the solution to the PB equation is changed,
is determined from
\begin{equation}
        \xi_c=
            \frac{\ln D/R_{ca}(\xi_c)}{1+\ln D/R_{ca}(\xi_c)},
\label{eq:thresh}
\end{equation}
in which $R_{ca}(\xi_c)$ is given by Eq. (\ref{eq:R}). 
Equation (\ref{eq:thresh}) is in fact obtained as an extension of
Eq. (\ref{eq:xicFuoss}) to the present model with constant volume
constraint \cite{Note1}.




\subsection{The chain elastic free energy}
\label{sub:elas}

We assume a freely-jointed-chain (FJC) 
model to calculate elastic contributions 
to the total free energy of polyelectrolyte chains in the brush. 
 The exact free energy of such a chain model 
has a purely entropic origin and is obtained (in units of $k_BT$) as 
\begin{equation}
    \frac{{\mathcal F}^{FJC}}{N}=
                  -\ln \frac{\sinh y}{y}
                  +y\coth y 
                  -1,
\label{eq:Elasticity}
\end{equation}
where $N$ is the number of monomers and $y$ is found from
\begin{equation}
    \frac{L}{L_0}=
                  \coth y -\frac{1}{y},
\label{eq:Elasticity2nd}
\end{equation}
in which $L$ and $L_0=N b_0$ are end-to-end distance and
contour length of the chain respectively (see Appendix \ref{app:A}).
In terms of the modeled polyelectrolyte rod, $L$ stands for the actual
height of the rod and $L_0$ 
for its height in the fully stretched situation. 
In the weak-stretching limit, 
Eqs. (\ref{eq:Elasticity}) and (\ref{eq:Elasticity2nd}) 
lead to the free energy of a Gaussian chain 
\begin{equation}
   \frac{{\mathcal F}^{FJC}}{N} 
              \approx 
                 \frac{3L^2}{2 (N b_0)^2}.
                 \label{eq:gausselas}
\end{equation} 
Whereas in the opposite limit of strong stretching, 
the model produces a non-linear elasticity with
\begin{equation} 
   \frac{{\mathcal F}^{FJC}}{N}
         \approx 
          -\ln(1-\frac{L}{Nb_0})+const.,
\label{eq:nonelas}
\end{equation}
which is dominated by the finite length of the chain.  
It is clear that for almost-fully-stretched chains, 
deviations from the Gaussian behavior will be important
\cite{PRY96}.




\subsection{Optimal brush height and its limiting behavior}
\label{sub:Osm}

The total free energy of the brush per unit cell 
is the sum of the electrostatic and elastic free energies 
obtained in Eqs. (\ref{eq:betafree}) or (\ref{eq:betafree1}) 
and (\ref{eq:Elasticity}),
\begin{equation}
   {\mathcal F}^{tot}=
                    {\mathcal F}^{PB}+{\mathcal F}^{FJC}.
\label{eq:freetot}
\end{equation}
The total free energy 
can be viewed as a function of the effective Manning 
parameter of the system, $\xi$, which varies according to the 
brush height $L$ (see Eq. (\ref{eq:xiL})). 
It also depends on the system parameters, 
namely, charge fraction of the rod, $f$, 
valency of charged monomers, $q$, and that of 
counterions, $q_c$, and finally on the ratio of the Bjerrum length, $\ell_B$, 
the cell radius, $D$, and the counterion size, $2r_c$, 
to the monomer size $b_0$.

\begin{figure}[t]
\begin{center}
\includegraphics[angle=-90,width=7.5cm]{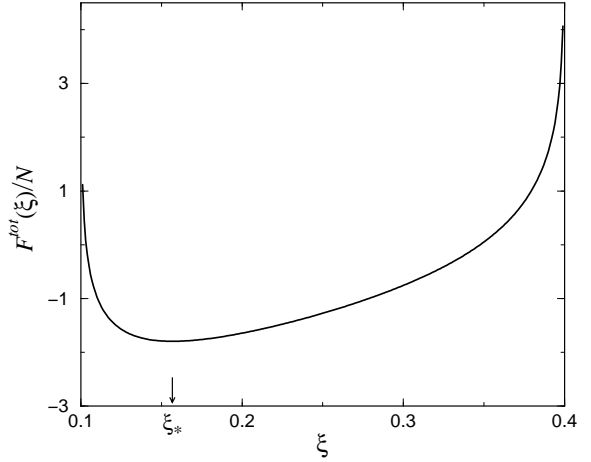}
\caption{\label{fig:free} Typical free energy of the present model
(per number of monomers) is plotted as a function of the Manning parameter
associated with the polyelectrolyte rod. Here it is calculated 
for  $f=1$, $\ell_B=0.1b_0$, $q_c=q=1$, $r_c=b_0/2$,
and $D=1.5b_0$. In this case, the lower and upper bounds 
for $\xi$ are $\xi_0=0.1$ and $\xi_u=0.4$ respectively. 
The arrow shows the location of the optimal Manning parameter
$\xi_*\approx 0.16$.  
} 
\end{center}
\end{figure}

The typical form of the total free energy (per unit cell)
as a function of $\xi$ is shown in Figure \ref{fig:free}, where the free energy
 is calculated for   
$f=1$, $\ell_B=0.1b_0$, $q_c=q=1$, $r_c=b_0/2$, and $D=1.5b_0$. 
As explained in Section \ref{sub:elec}, 
the effective Manning parameter, $\xi$, is bounded from below 
by $\xi_0$, and from above by $\xi_u$ given by Eq. (\ref{eq:upxi}).
In this example, the lower and upper bounds on $\xi$ are respectively
$\xi_0=0.1$ and $\xi_u=0.4$.
As seen in the graph, 
the total free energy has a minimum at an intermediate value of
$\xi$. The reason is that by decreasing 
$\xi$ down to $\xi_0$, the chain becomes highly stretched and its 
elastic free energy (see Eq. (\ref{eq:nonelas})) 
and consequently its total free energy 
increases and eventually diverges due to its finite extensibility.
In the other limit of large $\xi$,
the available space for counterions decreases as $\xi\rightarrow \xi_u$
(equivalently $R_{ca}\rightarrow D$);
thus the translational entropy of counterions 
dramatically decreases leading to an increasing free energy in this limit.
The optimal Manning parameter, $\xi_*$, that minimizes the total free 
energy, lies somewhere between these two limits
and corresponds to an optimal brush height $L_*=\xi_0 L_0/\xi_*$.

We thus obtain the optimal brush height, $L_*$, by minimizing the total free energy of the system 
(\ref{eq:freetot}) for various grafting densities. The results will be 
compared with the simulation data in Section \ref{sec:sim}. 
In this Section, we focus on
the generic predictions of the present model for the brush
height and its limiting behavior. To this end and 
for the sake of simplicity, we consider a simpler version of the model, 
in which the counterions are taken 
as point-like particles ($r_c=0$). Hence, as in Eq. (\ref{eq:oldR}),
\begin{equation}
   R=R_{ca}=R_0\sqrt{\xi/\xi_0}.
\label{eq:newR}
\end{equation}
We also take the charge valencies as $q_c=q=1$.

\begin{figure}[t]
\begin{center}
\includegraphics[angle=-90,width=7.5cm]{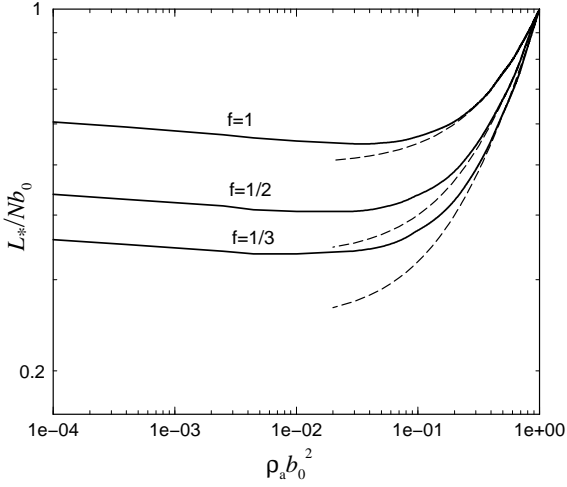}
\caption{\label{fig:limitlines1} Log-log plot of the rescaled 
optimal height of the brush 
as a function of the grafting density for   $\ell_B=0.1b_0$, $q_c=q=1$ and point-like counterions
(solid curves). 
Each of the curves corresponds to a charge fraction 
indicated on the graph.
Dashed lines are the asymptotic behaviors obeying Eq. (\ref{eq:asympt1})
for large grafting densities. 
} 
\end{center}
\end{figure}

Figure \ref{fig:limitlines1} shows the  
optimal brush height plotted as a function of grafting density 
for the Bjerrum length of $\ell_B=0.1b_0$, and for three different values 
of charge fraction $f=1$, 1/2 and 1/3 (solid curves). 
It is observed that the 
brush height has a non-monotonic behavior as a function of grafting density:
both in the limit of large grafting densities and small grafting 
densities, the brush height increases and eventually 
tends to its maximum value $L_0=Nb_0$.
Therefore, a lower bound is predicted for the equilibrium height 
of the brush that clearly depends on the charge fraction and 
the Bjerrum length. For charge fractions of $f>1/2$ and 
$\ell_B/b_0\sim 0.1$, this lower bound is about $50\%$ of the
contour length. This means that the polyelectrolyte rod remains 
increasingly stretched over a wide range of grafting densities.
The limiting behavior of the brush height with grafting density
can be understood both by asymptotic expansion of the free energy 
(see Appendix \ref{app:B}) and by simple physical arguments 
as we present now.

First, we consider the limit of large grafting densities 
$\rho_{\mathrm{a}}\rightarrow \rho_{\mathrm{a}}^{max}$, 
in which $\rho_{\mathrm{a}}^{max}=1/b_0^2$ 
is the maximum grafting density. This corresponds to the 
limit of small cell radius, {\em i.e.} $D/R_0\rightarrow 1$,
where $R_0=b_0/2$ in our discussion. 
In this limit, the available space for 
counterions inside the unit cell tends to zero and their 
entropy divergently decreases. 
The produced osmotic pressure becomes the major repulsive pressure
swelling the rod against stretching pressure of the elasticity, and bare
electrostatic effects become negligible. 
The entropy of counterions is then well-approximated by the entropy of
an ideal gas of particles, and can be written 
(per number of monomers and up to an additive constant) as 
\begin{equation}
   s_{ci}\approx 
              f\ln [\pi(D^2-R^2)L],
\label{eq:identropy}
\end{equation}
where $R$ is defined in Eq. (\ref{eq:newR}) (see Appendix \ref{app:B}). 
In this limit, the chain has a large extension and 
its elastic free energy per number of monomers is given by
Eq. (\ref{eq:nonelas}),
\begin{equation}
   f_{elas}\approx 
               -\ln(1-L/L_0).
\label{eq:nonelas1}
\end{equation}
Balancing the longitudinal pressures due to these
two opposing contributions,
\begin{equation}
  \frac{\partial}{\partial L} (f_{elas}-s_{ci})=0
\end{equation}
at fixed cell radius $D$, we find
\begin{equation}
        \frac{L_*(\rho_{\mathrm{a}})}{L_0}\approx
                       \frac{f+\rho_{\mathrm{a}} b_0^2}{1+f},
\label{eq:asympt1}
\end{equation}
for $\rho_{\mathrm{a}}b_0^2\rightarrow 1$, 
where, by the definition of the model in Section \ref{sec:model},
\begin{equation}
  \rho_{\mathrm{a}}b_0^2=
                      \frac{R_0^2}{D^2}.
\label{eq:rhoa}
\end{equation}
The expression given by Eq. (\ref{eq:asympt1})  is plotted in
in Figure \ref{fig:limitlines1} 
for the charge fractions $f=1$, 1/2, 1/3 (dashed curves), 
where the coincidence with the predictions from
minimizing the full free energy in Eq. ({\ref{eq:freetot}) (solid curves)
is seen at large grafting densities.

Note that within our model, the constant
volume constraint dominates when $\rho_{\mathrm{a}}b_0^2\rightarrow 1$,
and the linear dependence on the 
grafting density in Eq. (\ref{eq:asympt1}) is induced by this 
constraint. It is important to note
that the limiting behavior,  Eq. (\ref{eq:asympt1}), identifies the behavior of the brush height if
counterions around chains are taken as a {\em uniform} ideal gas
of particles with {\em no} lateral 
electrostatic effects (see Appendix \ref{app:B}). 
But, as clearly seen from Figure \ref{fig:limitlines1}, 
at lower grafting densities
({\em e.g.} at about $\rho_{\mathrm{a}}b_0^2\sim 0.1$, which
approximately 
corresponds to the simulated regime--see 
Figures \ref{fig:xi0p1} and \ref{fig:xi0p33} below), the predicted brush height
(solid curves) 
deviates from the above limiting line (dashed curves) displaying a weaker
dependence on the grafting density. 
This behavior at moderate grafting densities
is induced  by lateral
electrostatic effects, which become increasingly important and 
generate a minimum at intermediate 
grafting densities (Figure \ref{fig:limitlines1}). 

In the limit of small grafting densities 
$\rho_{\mathrm{a}}b_0^2\ll 1$, or equivalently 
$D/R_0\gg 1$, the present model is applicable for 
very long chains, since only in this case, counterions will be 
confined within the brush.
In fact, to retain
counterions inside the polyelectrolyte layer, 
one needs to consider chains of large charge fraction
with the number of monomers $N>(\rho_{\mathrm{a}}b_0^2)^{-1/2}$ 
(see the Discussion). 

In this limit, the
brush height shows different asymptotic behavior in terms  
of the grafting density depending on whether
the optimal Manning parameter, $\xi_*$, is below or above 
the threshold value $\xi_c$ as defined in Eq. (\ref{eq:thresh}). 
Thus, we shall distinguish two different 
scenarios. 
First we
look at the case when the optimal Manning parameter of the rod is smaller than the
threshold Manning parameter $\xi_*<\xi_c$. 
For our choice of parameters in Figure \ref{fig:limitlines1}
($\ell_B=0.1b_0, q=q_c=1$),  this holds  for $\rho_{\mathrm{a}}b_0^2 < 0.1$,
as can be checked from Eq. (\ref{eq:thresh}). 
In such conditions, the counterion cloud is highly diluted
(counterions effectively de-condense \cite{OOS71,MAN69,ZIMM}) 
and there will be no electrostatic
screening on the bare electrostatic potential of the rod 
$\psi(r)=2\xi\ln r$ \cite{NET98}.
This potential can be used to calculate the 
electrostatic energy (per number of monomers) 
\begin{equation}
   u_{elec}\approx 
               f\xi\ln \frac{D}{R},
\end{equation} 
where $R$ is defined in Eq. (\ref{eq:newR}) and $\xi$ is related to
the rod length $L$ through Eq. (\ref{eq:xiL}). 
The residual entropic contribution of counterions
may still be accounted for by assuming an ideal-gas entropy,
$s_{ci}$, as 
in Eq. (\ref{eq:identropy}). 
The electrostatic free energy of the system is then written as
${\mathcal F}_{elec}/N \approx u_{elec} - s_{ci}$ 
which, for very large $D/R_0$, may be approximated by
\begin{equation}
     \frac{{\mathcal F}_{elec}}{N} \approx 
                f(\xi-2) \ln \frac{D}{R}.
\label{eq:freeapp}
\end{equation}
This expression is
indeed a leading-order term and can be derived by expanding
the PB free energy (\ref{eq:betafree}) 
in terms of $R/D$-powers (see Appendix \ref{app:B}).
Using Eqs. (\ref{eq:freeapp}) and (\ref{eq:newR}), 
the longitudinal electrostatic pressure
can be calculated by differentiating ${\mathcal F}_{elec}$ with respect 
to $L=\xi_0L_0/\xi$.
Since the polyelectrolyte rod is again highly  
stretched in this limit, it is reasonable to approximate the  
elastic contribution by the same non-linear (logarithmic) expression,
Eq. (\ref{eq:nonelas1}), as found in the strong-stretching limit. 
The equilibrium brush height is obtained by balancing these two contributions to
the longitudinal pressure
on the polyelectrolyte rod ({\em i.e.} using Eqs. (\ref{eq:nonelas1}) and (\ref{eq:freeapp})).
The result (which has been calculated  numerically) 
is shown in Figure \ref{fig:limitlines2} 
(dashed curve) for $f=1$ and $\ell_B=0.1b_0$,  where
we also show the results from minimization of the full free energy,
Eq.  (\ref{eq:freetot}) (solid curve a).
The plot is made for $\rho_{\mathrm{a}}$ down to $10^{-6}b_0^{-2}$.
For vanishingly  small grafting densities $\rho_{\mathrm{a}}b_0^2\rightarrow 0$, 
the entropic contributions 
become negligible compared with the  
bare electrostatic repulsion between monomers, and
the equilibrium brush height behaves asymptotically as
\begin{equation}
    \frac{L_*(\rho_{\mathrm{a}})}{L_0}\approx
	                          \frac{f \ln \rho_{\mathrm{a}} b_0^2}
	                               {f \ln \rho_{\mathrm{a}} b_0^2-2\xi_0^{-1}}. 
\label{eq:asympt2below}
\end{equation}  
This function is not shown in Figure \ref{fig:limitlines2}, because it is valid
for smaller grafting densities. 
\begin{figure}[t]
\begin{center}
\includegraphics[angle=-90,width=7.5cm]{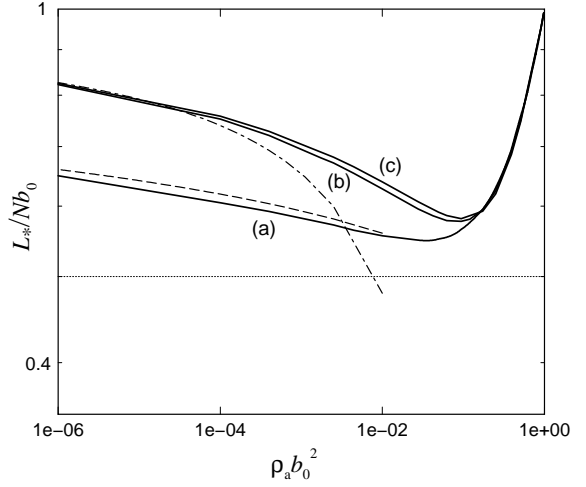}
\caption{\label{fig:limitlines2}  Log-log plot of the rescaled 
optimal height of the  
brush as a function of grafting density. Solid curves show the 
results  obtained from the minimization of the full free energy, Eq. (\ref{eq:freetot}), with
 $f=1$, $q_c=q=1$, and point-like counterions, for
a) $\ell_B=0.1b_0$ ($\xi_0=0.1$), b) $\ell_B=0.7b_0$ ($\xi_0=0.7$) 
and c) $\ell_B=1.2b_0$ ($\xi_0=1.2$).
The dashed and dot-dashed curves show the corresponding
asymptotic estimates at
small grafting densities for the cases a) (using Eq. (\ref{eq:freeapp})) and
c) (Eq. (\ref{eq:asympt2above})) respectively. The dotted line shows the brush height
 in the absence of lateral
effects for $f=1$ (Eq. (\ref{eq:nonlinear-scaling})). 
} 
\end{center}
\end{figure}

In the second scenario, {\em i.e.} 
when the optimal Manning parameter associated with the chains becomes  
larger than the threshold $\xi_*>\xi_c$, bare electrostatic
interactions are partially screened as a result of the
counterion-condensation process \cite{OOS71,MAN69,ZIMM}. 
In fact, the PB electrostatic potential  
(up to some logarithmic corrections) 
reduces to the bare electrostatic potential of 
a rod with critical Manning parameter
$\xi_M=1$, {\em i.e.}, $\psi(r)=2\ln r$, when 
$D/R_0\rightarrow\infty$ \cite{Note1,NET98}.
The electrostatic energy of the system for $D/R_0\gg 1$ can be 
estimated using this potential, which yields
\begin{equation}
   u_{elec}\approx 
                  \frac{f}{\xi}\ln \frac{D}{R}.
\label{eq:ularge}
\end{equation}
To estimate entropic contributions in this case,
we may adopt the counterion-condensation picture \cite{OOS71,MAN69}
that only a fraction of $1/\xi$ of counterions are unbound and may 
contribute to the entropic pressure. Thus,
the corresponding 
ideal-gas entropy of counterions, $s_{ci}$ in Eq. (\ref{eq:identropy}), may
be corrected by such a factor 
and used, together with Eq. (\ref{eq:ularge}), to derive the 
leading term of the electrostatic free energy, 
${\mathcal F}_{elec}/N \approx u_{elec} - s_{ci}$, as
\begin{equation}
     \frac{{\mathcal F}_{elec}}{N} \approx 
                -\frac{f}{\xi} \ln \frac{D}{R}.
\label{eq:freeapp_above}
\end{equation}  
The expression (\ref{eq:freeapp_above}) is 
confirmed again by a limit expansion of the PB free energy  
(\ref{eq:betafree1})--see Appendix \ref{app:B}. Calculating the longitudinal 
electrostatic pressure from Eq.
(\ref{eq:freeapp_above}) and balancing it with the 
non-linear stretching pressure from Eq. (\ref{eq:nonelas1}), 
one finds that
\begin{equation}
    \frac{L_*(\rho_{\mathrm{a}})}{L_0}\approx
                 1+\frac{2\xi_0}
                        {f\ln \rho_{\mathrm{a}} b_0^2}.
\label{eq:asympt2above}
\end{equation} 
The asymptotic expression, Eq. (\ref{eq:asympt2above}),  
is shown in Figure \ref{fig:limitlines2}
(dot-dashed curve) along with the result from minimization 
of the full free energy, Eq.  ({\ref{eq:freetot})  (solid curve c) for a system with
 $f=1$ and $\ell_B=1.2b_0$ ($\xi_0=1.2$)
for which the optimal 
Manning parameter, $\xi_*$,
remains always above the threshold \cite{Note1}.  
The above estimate, Eq. (\ref{eq:asympt2above}), represents the result  
obtained from the model for $\rho_{\mathrm{a}}b_0^2 < 10^{-4}$. 
At small grafting densities, the 
constant volume constraint becomes unimportant.
We also note that within our model, 
similar behavior is obtained for the
brush height
for the whole range of Bjerrum lengths, see the result for 
$f=1$ and $\ell_B=0.7b_0$ 
($\xi_0=0.7$) in Figure \ref{fig:limitlines2} (solid curve b).

The preceding discussion on the limiting behavior of the brush 
thickness demonstrates the important role of lateral electrostatic 
contributions
to the total free energy that generate re-stretching of the chains 
at small grafting densities. 
Also we showed that the strength of these effects is controlled by the 
counterion-condensation process, which 
is systematically included
in the PB equation used in Section \ref{sub:elec} \cite{HOL01,ZIMM,NET98}.
Both for weakly charged ($\xi_0<1$) and highly charged chains ($\xi_0>1$), 
lateral electrostatic contributions produce a repulsive longitudinal force acting
on the chains, which logarithmically 
increases by decreasing the grafting density (Eqs. (\ref{eq:freeapp}) 
and (\ref{eq:freeapp_above})). In particular, 
for highly charged chains, the force is independent
of the brush height, {\em i.e.} 
$-\partial {\mathcal F}_{elec}/\partial L\sim \ln D/R$, and results 
from the electrostatic screening 
due to condensation of counterions 
(see Eqs. (\ref{eq:freeapp_above}) and (\ref{eq:xiL})).
In any case, the increase of the brush height, which converges  
to the contour length, 
is logarithmically weak as the grafting density is lowered.
We note that this behavior will not be obtained if 
lateral effects are neglected. In this case, the brush height
remains independent of the grafting density, 
\begin{equation}
    \frac{L_*}{L_0}=\frac{f}{1+f},
\label{eq:nonlinear-scaling}
\end{equation} 
which is shown by a dotted line in Figure \ref{fig:limitlines2}, and 
follows from balancing the entropic term, Eq. (\ref{eq:identropy}), 
and the non-linear elasticity, Eq. (\ref{eq:nonelas1}), where 
 the volume constraint, Eq. (\ref{eq:newR}), is also neglected

As already mentioned, our model  describes the situation, in which 
counterions remain all inside the brush
and distribute uniformly in the direction normal to the anchoring plane. 
In the range of parameters, 
when a considerable fraction of counterions  
leaves the brush ({\em e.g.} when chains are short or charge fraction is small), 
a different behavior will be obtained for
the brush thickness as a result of the inhomogeneous distribution
of counterions in the normal direction. In this case, if the concentration of
counterions (and that of monomers) is assumed to be smeared out laterally, 
the brush height is obtained to decrease 
monotonically by lowering the grafting density as shown in Ref. \cite{ZHU97}. 
Therefore, the behavior of the brush height in the presence of 
non-uniformities in both lateral and normal directions 
should be examined in an extended approach. 

Finally, we remark that the non-monotonic behavior of the 
brush thickness is not
influenced by the elasticity model and 
qualitatively similar features are obtained when a Gaussian 
chain elasticity is used \cite{Note_elas}.




\section{Comparison to molecular dynamics simulations}
\label{sec:sim}

Computer simulations provide an excellent mean to study polymer
systems. Extensive molecular dynamics simulations have been
performed recently on polyelectrolyte brushes at 
various grafting densities and charge fractions, both at strong
\cite{CSA01,CSA99,CSA00} and intermediate
\cite{CS02} electrostatic couplings.

In these simulations, a freely-jointed bead-chain model is adopted where the 
monomers are connected by non-linear springs with the so-called
FENE (finite extensible non-linear elastic) potential, 
and end-grafted onto a rigid 
surface. The counterions are explicitly modeled as charged 
particles and no additional electrolyte is added. The simulation box 
is periodic in lateral directions and finite in the $z$-direction normal 
to the anchoring surface at $z=0$. We use the techniques introduced in Refs. 
\cite{SPE,ARNOLD} 
to account for the long-range nature of 
the Coulombic interactions in a laterally periodic system.  The short-range 
repulsion between particles separated by distance $r$ is modeled by a shifted 
Lennard-Jones (LJ) potential 
$u(r)=4\epsilon_{LJ}\{(\sigma/r)^{12}-(\sigma/r)^{6}+1/4\}$ 
for $r/\sigma \leq 2^{1/6}$ and 
$u(r)=0$ otherwise, with Lennard-Jones diameter of $\sigma$
being equal for both monomers and counterions.
The counterions and charged monomers are univalent ($q_c=q=1$).  
For the simulations reported below, we choose moderate values for 
the Bjerrum length, {\em i.e.} $\ell_B=\sigma$ and $\ell_B\approx 2\sigma$, 
which correspond to  
electrostatically-intermediate-coupling situation. 
With this choice of parameters, the average 
bond length (which is the result of the interplay between LJ repulsion
and the FENE bond potential) is almost unaffected by the electrostatic 
repulsions and is $b_0=0.98\sigma$.

\begin{figure}[t]
\begin{center}
\includegraphics[angle=0,width=3.5cm]{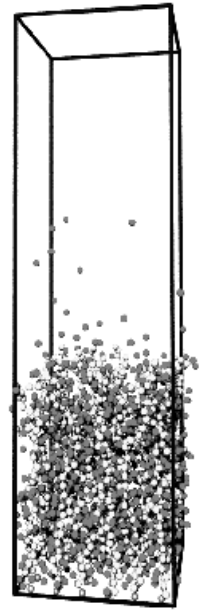}
\caption{\label{fig:snapshot}  A snapshot of the simulated 
polyelectrolyte brush with 36 fully charged 
chains of $N=30$ monomers (in light gray) at 
grafting density of $\rho_{\mathrm{a}}=0.12 \sigma^{-2 }$.  
The Bjerrum length is $\ell_B=\sigma$. Counterions are shown by 
dark gray spheres. The box height perpendicular
to the anchoring plane has been reduced for the
sake of representation.}
\end{center}
\end{figure}

\begin{figure}[t]
\begin{center}
\includegraphics[angle=0,width=9.5cm]{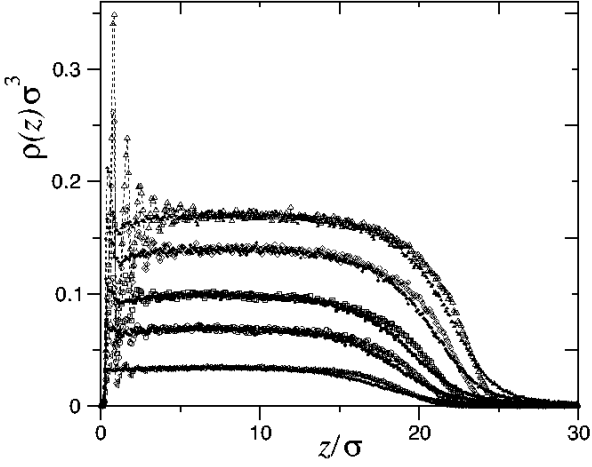}
\caption{\label{fig:Profiles} Density profiles of monomers 
$\rho_m(z)$ (open symbols) and
counterions $\rho_{ci}(z)$ (filled symbols) as a function of the
distance from the anchoring surface. Shown are profiles for 
fully charged brushes of
36 chains of $N=30$ monomers at $\ell_B=\sigma$ and 
grafting densities (from bottom
to top) $\rho_{\mathrm{a}}\sigma^2 =0.020$ (triangle-lefts), $0.042$
(circles),   $0.063$ (squares), $0.094$
(diamonds), and $0.12$ (triangle-ups).} 
\end{center}
\end{figure}

Figure \ref{fig:snapshot} shows a snapshot from the simulation of a brush
with 36 chains of 30 monomers, which is fully
charged and has a large grafting density of 
$\rho_{\mathrm{a}}=0.12 \sigma^{-2 }$. 
In this figure,
the connectivity of the chains are preserved, so that the chains may 
appear to extend 
beyond the simulation box. (The simulation image has been 
produced using the VMD software package \cite{VMD}.) 
Simulated density profile of monomers and 
counterions of the system in normal direction are shown in Figure 
\ref{fig:Profiles} for the fully charged brush
at several grafting densities.  
As seen, both monomers and counterions 
follow very similar 
nearly-step-like profiles with uniform amplitude inside the
brush, which increases with grafting density (the monomers show a short-range ordering 
close to the anchoring plane, which 
is not relevant in the present study).
These figures 
show that the counterions are mostly confined in the brush layer
and that the electroneutrality condition is satisfied locally \cite{CS02}.
One may observe (more clearly from simulated brush heights
in Figures \ref{fig:xi0p1} and \ref{fig:xi0p33}) that the
polyelectrolyte chains are stretched up
to about 60$\%$ of their contour length, and thus their
elastic behavior is far beyond the linear regime. 
Therefore, within the chosen range of parameters, the
simulated brush is in the strong-charging and strong-stretching
limits and as we will show, it 
exhibits the non-linear osmotic brush regime.

\begin{figure}[t]
\begin{center}
\includegraphics[angle=0,width=8.cm]{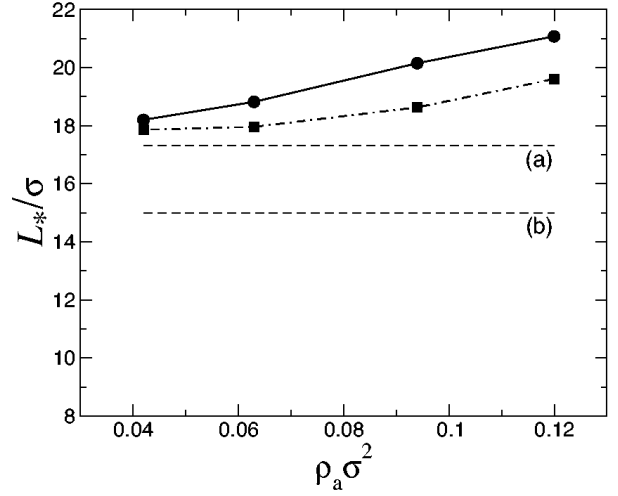}
\caption{\label{fig:xi0p1} Brush height as a function of grafting density for 
polyelectrolyte chains of $N=30$ monomers 
(contour length $L_0=30\sigma$) with charge fraction of $f=1$. Circles show the
simulation data and squares are the predictions of the present mean-field model. 
The dotted lines (a) and (b) show the scaling predictions, 
Eqs. (\ref{eq:Losmotic}) and (\ref{eq:nonlinear-scaling}), 
with Gaussian and non-linear elasticity  respectively. 
Here Bjerrum length is $\ell_B=\sigma$ and charged particles are univalent.
} 
\end{center}
\end{figure}

\begin{figure}[t]
\begin{center}
\includegraphics[angle=0,width=8.cm]{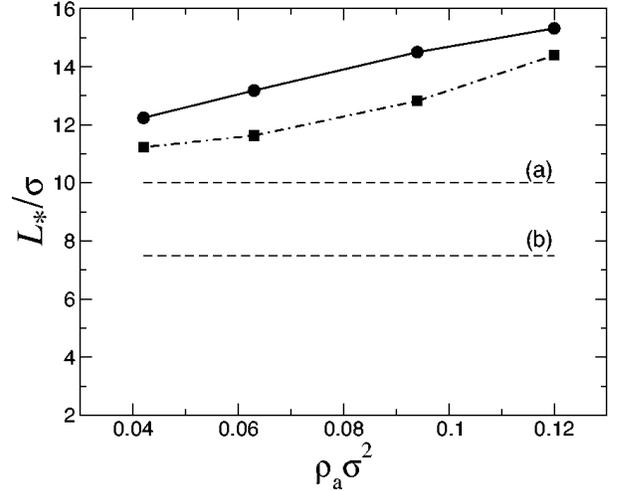}
\caption{\label{fig:xi0p33} Same as Figure \ref{fig:xi0p1} but for $f=1/3$ and 
Bjerrum length $\ell_B\approx 2 \sigma$.
} 
\end{center}
\end{figure}


The average height of endpoints of the chains is one of
the quantities which can be directly 
measured in the simulations. We compare the
predictions of our theoretical model for the brush height with the
average height of the endpoints from our simulations in Figures 
\ref{fig:xi0p1}  and \ref{fig:xi0p33} for charge fractions $f=1$ (with $\ell_B=\sigma$) and $f=1/3$ 
(with $\ell_B\approx 2\sigma$) respectively.
To obtain theoretical data from the model, 
we have taken into account the finite size 
of the counterions using $r_c=b_0/2$; moreover, we take $b_0=\sigma$
while comparing these data with the simulation results.

As seen from the Figures, the simulated brush height (circles)
varies slowly with the grafting density, contrary to the 
predictions of scaling theories. 
According to the scaling studies \cite{PIN91,BOR91,WIT93}, 
the osmotic brush thickness
is obtained as 
\begin{equation}
   L_{OsB}=
           \frac{1}{\sqrt{3}}Nf^{1/2}b_0,
\label{eq:Losmotic}
\end{equation} 
which is independent of the grafting density (shown by dotted line (a) in the Figures).
This relation is based on a balance between the osmotic pressure 
of an ideal gas of counterions inside the brush
$\pi_{os}=Nf\rho_{\mathrm{a}}/L$, and the elasticity of
Gaussian chains, $\pi_{elas}=3\rho_{\mathrm{a}}L/Nb_0^2$.
If we adopt the non-linear elasticity for large stretching, Eq. (\ref{eq:nonelas1}), the brush height 
is given by  Eq. (\ref{eq:nonlinear-scaling}), when lateral effects are neglected. 
The result is also independent of the grafting density  
and shown by dotted line (b) in Figures \ref{fig:xi0p1} and \ref{fig:xi0p33}.
The present model considers  corrections to
the scaling theory by accounting for lateral electrostatic and coupling effects.
The theoretical predictions (squares) 
appear to be in reasonable agreement with the simulation results. 
As shown in Ref. \cite{Naji03} by comparing different cell models for the brush,
the deviations from the simulation data are in fact of the oder of the systematic 
error of the employed cylindrical cell model  (Section \ref{sec:model}). 
In other words, choosing a different cell
volume or a different rod model for the polymer chain leads to 
differences in theoretical predictions 
that are of the same order as the deviations from the simulation data
in Figures  \ref{fig:xi0p1} and \ref{fig:xi0p33}.
Yet some of the differences may be in general 
due to excluded-volume effects 
that  have been treated in a simplified 
way in our study. In particular, we 
have neglected the excluded-volume repulsion between counterions; a 
rough estimate of these effects, however, shows that
such volume effects are almost one order of magnitude
smaller than the osmotic pressure of counterions (see the Discussion).

Another effect which may play a role in the simulated systems is 
lateral wiggling of the polyelectrolyte chains. In fact, the volume constraint,
that has been used in our model, can account for some of
these effects in a very simple way, since the effective chain radius
increases for decreasing brush height. However, conformational changes of the chains
may affect the validity of the cell model that was employed 
for calculating the electrostatic contributions to the free energy. 
From lateral monomer density profile obtained within the simulations \cite{CS_unpub},
we conclude 
that  wandering of chains is
well-confined to a cylindrical region, the size of which is smaller than the
``cell radius''. The cell radius is close to the decay
 range of the lateral 
monomer density profile only for large grafting densities, thus 
in this regime, lateral
wiggling of the chains might be responsible for some of
the deviations between the theoretical predictions and the simulation data.




\section{Conclusion and Discussion}
\label{sec:Discussion}

We have studied strongly charged polyelectrolyte brushes
in the non-linear osmotic regime. In order to bring out the features of the 
non-linear osmotic brush regime more clearly, we chose 
a moderate Bjerrum length of the order of the monomer size in our simulations.
Nonetheless, the present theory is obtained for the whole range of Bjerrum lengths, 
and similar features are obtained 
at large, more realistic values of the Bjerrum length 
(see Section \ref{sub:Osm}).

In the simulations \cite{CS02}, we
have observed that
the brush height varies slowly 
with the grafting density of the brush. In this
situation, polyelectrolyte chains are strongly stretched 
and counterions are mostly confined within the brush layer.
 In the direction normal to
the anchoring plane, 
counterions produce an almost step-like density profile, 
while in lateral directions, an inhomogeneous
distribution is observed around polyelectrolyte chains 
at moderate grafting densities  \cite{CSA99,CSA00,CS02}.

We have investigated this system using a theoretical model, in which 
we consider a number of corrections to the standard scaling
theories. Firstly, the 
electrostatic free energy of the system is calculated within
the non-linear Poisson-Boltzmann theory, which includes 
lateral electrostatic effects. On the other hand, 
the elastic free energy
of the chains is calculated using a freely-jointed-chain 
model providing  a non-linear elasticity in the strong-stretching
limit, which accounts for the finite extensibility of the 
brush chains. We have taken into account also the back influence
of the conformational changes of the chains on the osmotic pressure of counterions.
Such a coupling
between longitudinal and lateral degrees of freedom is modeled in a 
simple way by assuming a constant effective volume 
for the chains. 
The results of our model display
a non-monotonic behavior for the brush height as a 
function of the grafting density. 
In  particular, 
at moderate  grafting densities,
we find that the brush height weakly increases
with grafting density, which arises as a result of the interplay between
lateral electrostatics and the coupling (volume constraint) effect. 
This behavior agrees with the simulation data within the systematic error of our model;
the deviations may originate mainly from non-uniform boundary conditions 
in the simulations. Predictions of our model also agree with 
recent experiments on osmotic brushes on a semi-quantitative level
as discussed in Refs. \cite{HELM,GUENOUN}. 

As already discussed, the behavior of
the brush height at very large grafting densities (closely-packed
brush) is dominated by the volume 
constraint used in our model.
The characteristic of this regime is the linear increase of
the brush height with grafting density (see Eq. (\ref{eq:asympt1})). 
In this regime, excluded-volume effects become relevant and 
should be considered explicitly in the balance equations. While, in our study,
volume interactions are only effectively modeled.
Nonetheless, at lower grafting densities ({\em e.g.} at about 
$\rho_{\mathrm{a}} b_0^2\sim 0.1$ as investigated in the simulations), 
lateral electrostatic effects involved in the PB
free energy become increasingly important and lead to a weaker 
increase of the brush height with grafting density (see Figure 
\ref{fig:limitlines1}).  

For large grafting densities, possible contribution 
of the excluded-volume repulsions between counterions may be estimated 
as follows. Effectively, 
the second-virial contribution from counterions to the total free energy  
(per unit cell and in units of $k_BT$) may be written as 
\begin{equation}
  \frac{{\mathcal F}_{v2}}{N}=\frac{1}{2N}v_2 \,\overline{\rho}^2_{ci}V_{cell}=
        \frac{1}{2}v_2\frac{Nf^2}{\pi (D^2-R^2)L},
\label{eq:v2}
\end{equation}
where $v_2>0$  is the effective virial coefficient (for 
good solvent condition), and 
$\overline{\rho}_{ci}$ is the average density of counterions
in a unit cell of volume $V_{cell}=\pi (D^2-R^2)L$. 
The longitudinal pressure coming from the excluded-volume repulsions 
between counterions, $\pi^{Long}_{v2}$, is calculated by differentiating 
Eq. (\ref{eq:v2}) with respect to $L$, the brush height. The 
produced pressure is then compared with the longitudinal osmotic pressure 
of counterions obtained from Eq. (\ref{eq:identropy}), 
that is $\pi^{Long}_{os}\sim f/L$.
We find
\begin{equation}
  \frac{\pi^{Long}_{v2}}{\pi^{Long}_{os}}\sim 
                  \frac{v_2/2\pi b_0^3}{L/Nfb_0}
                   \left(\frac{4\rho_{\mathrm{a}} b_0^2}
                         {1-4\rho_{\mathrm{a}} b_0^2}\right),
\end{equation}
where we have also considered the closest approach distance between counterions 
and monomers as discussed in the text (here we have taken 
univalent particles for simplification). 
Now, for largest grafting densities used in the simulations, 
{\em e.g.} $\rho_{\mathrm{a}} b_0^2\sim 0.1$,
and for fully charged chains $f=1$ (where $L/L_0\sim 2/3$), 
it follows that 
${\pi^{Long}_{v2}}/{\pi^{Long}_{os}}\sim  (1/10)v_2/v_2^{HC}$, where
$v_2^{HC}=\pi b_0^3/6$ is the hard-core second virial coefficient. This 
ratio is even smaller for smaller charge fractions and grafting densities,
and roughly yields the largest estimate based on the simulation data. 
Clearly, if we assume that $v_2\sim v_2^{HC}$, the
excluded-volume pressure of counterions, which is neglected in our
model, is found to be almost one order 
of magnitude smaller than their osmotic pressure.

At small grafting densities, our model predicts a weak re-stretching of the
chains by lowering the grafting density, which is produced by lateral electrostatic
contributions and is controlled by
lateral rearrangement of counterions (counterion-condensation process)
around the chains (Section \ref{sub:Osm}). These results are valid 
only if the optimal height of the brush is much
larger than the distance between neighboring grafted chains. 
(In particular, the
counterion condensation occurs in the presence of long chains, when 
end effects are negligible.) This situation may be achieved 
by taking  polyelectrolyte chains with sufficiently large 
contour length and at large charge fractions.
Otherwise, the cylindrical symmetry, assumed in the 
calculation of the electrostatic free energy, and also the 
osmotic conditions are  no longer 
retained. We may estimate the regime of validity of our results 
as follows. At small grafting densities, 
a large fraction of counterions tends to leave the brush
and consequently, the
system is pushed away from the osmotic regime: for large charge
fractions, it will enter the Pincus brush regime,
where uncompensated electrostatic repulsion between monomers 
balances the elastic stretching, 
and for small charge fractions of the chains, 
mushroom conformations, that are entropically favorable,
are formed by shrinkage of the chains onto the anchoring plane. 
The crossover from osmotic to Pincus brush regime can be
identified by introducing the average height 
of the counterion layer over the anchoring plane.
The thickness of the counterion layer is in fact
characterized by the effective Gouy-Chapman length of the
plane, that is   
$\lambda_{GC}=1/2\pi \ell_B N f \rho_{\mathrm{a}}$ (we
restrict our discussion to the case with univalent particles).
In the osmotic brush regime, where the surface charge density 
$N f \rho_{\mathrm{a}}$ is large, one has
$\lambda_{GC}\leq L$, whereas in 
Pincus brush regime with small surface charge density 
$\lambda_{GC}\gg L$. (In the simulations, we have 
$\lambda_{GC}/ L\sim 10^{-3}$ for $f\sim 1$ and 
$\rho_{\mathrm{a}} b_0^2\sim 0.1$.)
To retain counterions inside the brush, one needs 
to take polyelectrolyte chains with number of monomers $N$
larger than a certain threshold  number  $N_*$. 
To estimate $N_*$, one can use scaling arguments 
 that yield Pincus brush height as \cite{PIN91}
\begin{equation}
    L_{PB}\sim 
              N^3\ell_B f^2\rho_{\mathrm{a}} b_0^2.
\label{eq:Lpincus}
\end{equation} 
Now, given the scaling relations for the osmotic and Pincus
brush regimes, Eqs. 
(\ref{eq:Losmotic}) and (\ref{eq:Lpincus}) respectively, 
one can find the boundary relation between these two regimes as 
$N^2f^{3/2}\ell_B \rho_{\mathrm{a}}b_0\sim 1$ \cite{BOR94}.
The threshold number of monomers $N_*$ follows from this relation, {\em
i.e.}, 
\begin{equation}
    N_*=
        f^{-3/4}(\rho_{\mathrm{a}}b_0^2)^{-1/2}
        (\frac{b_0}{\ell_B})^{1/2}.
\label{eq:Nstar1}
\end{equation}
One notes that for $f\sim 1$ and $\ell_B\sim b_0$ 
this corresponds to the overlapping threshold for
neighboring chains, that is when
$\rho_{\mathrm{a}} L_0^2\sim 1$ with $L_0=Nb_0$ as the contour length
of the chains. 
To give a numerical estimate of $N_*$, we use the simulation parameter
$\ell_B=b_0$ for fully charged chains ($f=1$): at grafting densities 
as small as $\rho_{\mathrm{a}}b_0^2\sim 10^{-6}$, Eq. (\ref{eq:Nstar1}) 
gives $N_*\sim 10^3$,
while for $\rho_{\mathrm{a}}b_0^2\sim 10^{-2}$ we would have 
$N_*\approx 10$. Note that the latter case coincides with the minimum
grafting density used in the simulations (see Section \ref{sec:sim}), where 
polyelectrolyte chains bear $N=30$ monomers.

Now, to prevent formation of mushrooms at small grafting densities
and {\em small charge fractions}, one may
take longer chains with $N>N_{**}=(\rho_{\mathrm{a}}b_0^2)^{-1}$, which
is a more stringent condition on $N$, {\em i.e.} $N_{**}>N_*$. This
corresponds 
to the condition that the Gaussian size of the polyelectrolyte chains,  
$N^{1/2}b_0$, be larger than the distance between neighboring chains 
$\rho_{\mathrm{a}}^{-1/2}$. (Note that we have assumed a 
Gaussian polymer, thus 
our estimate gives an upper bound for $N_{**}$.) At small grafting density of 
$\rho_{\mathrm{a}}b_0^2\sim 10^{-6}$, one has
$N_{**}\sim 10^6$, a quite large value, and for 
$\rho_{\mathrm{a}}b_0^2\sim 10^{-2}$, we obtain $N_{**}\sim 100$.
We conclude that the regime, where the brush height has been predicted 
to increase with decreasing the grafting density is observable in experiments
and simulations by choosing long enough chains. 

In this paper the system is studied within a mean-field approximation, which is
valid only if electrostatic correlations are negligible.
Such effects become important at strong electrostatic couplings \cite{NET01},
{\em i.e.} at large Bjerrum lengths and charge valencies, and can produce
significant attractive pressure on the chains.
In such conditions, a scaling theory has been developed 
which predicts a collapsed brush regime observed
also in simulation \cite{CSA01}.
At moderate Bjerrum length and for univalent particles 
as chosen in our simulations, 
correlation effects are not yet important, and as we showed,
the predictions of the present mean-field theory are very close to the 
simulation results.  

An interesting problem is to extend the present results
to include the variation of the counterionic density profile in the direction
normal to the anchoring plane, which may be caused by
the diffusion of counterions outside the brush at low grafting densities or at small
charge fractions. The self-consistent field analysis 
of Zhulina and Borisov \cite{ZHU97} reveals that if these effects are 
accounted for using the non-linear PB equation, the brush height
monotonically decreases by lowering the grafting density provided that the 
concentration of counterions (and that of monomers)
is assumed to be smeared out laterally. Our results show a re-increase 
of the brush height by lowering the grafting density, when the counterion
profile is assumed to be uniform in normal direction, and allowed to
admit a laterally inhomogeneous form according to the non-linear 
PB equation. An extended approach should, therefore, 
examine the interplay between these two mechanisms.

Finally, we remark that similar 
calculations can be done for slightly different models \cite{Naji03}, 
{\em e.g.} one may consider a volume charge distribution
for polyelectrolyte chains. Clearly, the Coulomb self-energy of the
polymer  would be larger in this case  and a larger brush height is
predicted as compared with the studied model.




\begin{acknowledgement}

We acknowledge discussions with O. Borisov and N.A. Kumar. 
We also thank H. Ahrens, C.A. Helm, and 
G. Romet-Lemonne, J. Daillant, and P. Guenoun for communication about
their recent experimental data. 
A.N. acknowledges financial support from the DFG Schwerpunkt Polyelektrolytes
with defined architecture and the DFG German-French Network. 
C.S. gratefully acknowledges grants for computer time at the John von
Neumann-Institut for Computing (NIC) J\"ulich and the Konrad Zuse
Zentrum f\"ur Informationstechnik Berlin (ZIB).

\end{acknowledgement}


 

\appendix




\section{The free energy of a freely-jointed chain}
\label{app:A}

We calculate the free energy of a freely-jointed chain consisting
of $N$ links (monomers), each of fixed length $b_0$ and able to point 
in any direction independently of each other.  It is convenient to perform 
the calculation in an isobaric ensemble, in which the chain is 
considered to be stretched 
by applying a constant force $F$ (in units of $k_BT$). 
Since the monomers are 
assumed as rigid links, the configuration space of the chain is spanned by 
a set of angles $\{\theta_i, \phi_i\}$ specifying orientations of the 
monomers
labeled by $i=1,\ldots, N$ (a standard spherical frame of coordinates
is chosen with  $z$-axis pointing in the same direction as the force). 
The partition function of such a polymer chain can be written as
\begin{eqnarray}
   {\mathcal Z}_F=\left[                                                
                        \int_0^{2\pi}
                        \frac{d\phi}{2\pi}              
                        \int_0^{\pi}
                        \frac{d\theta}{2}
                        \sin\theta \, e^{b_0F\cos\theta}                
                  \right]^{N}.
\label{eq:fjpZ}
\end{eqnarray}
Integrating  Eq. (\ref{eq:fjpZ}) we find
\begin{equation}
   {\mathcal Z}_F= \left[               
                        \frac{  e^{b_0F}- e^{-b_0F}}{2b_0F}             
                   \right]^{N}.
\end{equation}
Now, the extension (end-to-end distance) of the chain, $L$,  
is calculated from 
$L=\partial \ln {\mathcal Z}_F/\partial F$ as
\begin{equation}
    \frac{L}{b_0N}=                                                     
                \coth b_0F-\frac{1}{b_0F}.
\label{eq:exten}
\end{equation}
Using Legendre transformation we can calculate the isochoric
free energy of the system (in units of $k_BT$),
${\mathcal F}^{FJC}=-\ln {\mathcal Z}_F+LF$, hence
\begin{equation}
  \frac{{\mathcal F}^{FJC}}{N}=         
                  -\ln \frac{\sinh b_0F}{b_0F}                          
                  +b_0F\coth b_0F-1,
\label{eq:fjpF}                                                         
\end{equation}
which was used together with Eq.
(\ref{eq:exten}) in Section \ref{sub:elas}.

In the weak-stretching or Gaussian-chain limit $b_0F\ll 1$, the 
end-to-end 
distance is obtained by a proper expansion of  Eq. (\ref{eq:exten}) 
as $L\approx Nb_0^2F/3$, which is then used to get the free energy 
${\mathcal F}^{FJC}/N\approx 3L^2/2(Nb_0)^2$. In the strong-stretching limit 
$b_0F\gg 1$, a non-linear 
force-extension relation is reached from  Eq. (\ref{eq:exten}), that is 
$L/b_0N\approx 1-1/b_0F$. Using this, we find the elastic 
free energy ${\mathcal F}^{FJC}/N\approx -\ln (1-L/Nb_0)+const$. 




\section{Asymptotic expansions of the PB free energy}
\label{app:B}

Here, we briefly discuss the asymptotic expansions of the
PB free energy both in the limit of 
small and large grafting densities. For simplicity, 
we assume $q_c=q$.


\subsection{Small grafting densities ($D/R\gg 1$):}

In this case, there are different limit expansions
for the PB free energy  depending on whether the 
Manning parameter is smaller or larger than $\xi_c$.
Note that in this limit, the threshold Manning parameter itself
tends to one, {\em i.e.} $\xi_c\rightarrow 1$ 
(see Eq. (\ref{eq:thresh}), also Ref. \cite{FUOSS}).

\subsubsection{Case with $\xi\leq \xi_c$.}

Inspecting Eq. (\ref{eq:beta}), one can find that in the limit of
 $D/R\rightarrow \infty$, the solution for $\beta$ tends asymptotically
to $\beta=(\xi-1)$.
Hence for $D/R\gg 1$, we can propose the following form for $\beta$,
\begin{equation}
    \beta^2  \approx 
                    (\xi-1)^2(1-x), 
\label{eq:betax}
\end{equation} 
where $x$ is a small function of $\xi$ and $D/R$, and may 
be determined as follows.
Rearranging Eq. (\ref{eq:beta}), we can find a more convenient
 equation for
the forthcoming limit expansions,  
\begin{equation}
    \beta \ln \frac{D}{R}=
                      \frac{1}{2}\ln 
                                \frac{1-\beta}{1+\beta}
                      - \frac{1}{2}\ln 
                                \frac{(\xi-1)+\beta}{(\xi-1)-\beta}.
\label{eq:betanew}
\end{equation}
Now using Eq. (\ref{eq:betax}) into Eq. (\ref{eq:betanew}) and expanding for 
small $x$, we have
\begin{equation}
         x \approx 
                \frac{4\xi}{2-\xi}
                \left(
                        \frac{D}{R}
                \right)
                        ^{2(\xi-1)}.
\label{eq:x}
\end{equation}
Similar limit expansions can be performed for the potential field, 
Eq. (\ref{eq:Solution}), 
and the PB free energy, Eq. (\ref{eq:betafree}), with $x$ given
by Eq. (\ref{eq:x}), which lead to
\begin{equation}
        \psi(r)\approx 
                     2\xi\ln \frac{r}{R},
\end{equation}
(where we have assumed $\psi(R)=0$), and
\begin{equation}
       \frac{{\mathcal F}^{PB}}{N}
                                \approx 
                                f(\xi-2) \ln \frac{D}{R},
\label{eq:freeapp1}
\end{equation}
as estimated in Eq. (\ref{eq:freeapp}) in the text. 
Clearly, both bare electrostatic repulsions of the charged rod and 
entropy of mobile counterions  contribute
in Eq. (\ref{eq:freeapp1}) respectively as $f\xi \ln D/R$ and
$-2f \ln D/R$. However,
in the limit of very large $D/R$, the longitudinal entropic pressure of 
counterions becomes vanishingly small relative to the longitudinal 
bare electrostatic pressure.
This can be seen from Eq. (\ref{eq:freeapp1}),
noting that
the longitudinal pressure
is obtained by differentiation of the corresponding term in the free energy
with respect to the cell volume $\pi D^2 L$ at fixed cell radius $D$. Formally,
the longitudinal osmotic pressure, $\pi_{os}^{Long}$, will be
$\pi_{os}^{Long}=-\partial{\mathcal F}_{elec}/\pi D^2\partial L$,
where $L$ is related to $\xi$ via Eq. (\ref{eq:xiL})
and ${\mathcal F}_{elec}$ is given by Eq. (\ref{eq:freeapp1}).

In contrast, both bare electrostatic repulsions and entropic
effects produce lateral pressures of the same leading order.
(The lateral pressure is obtained by differentiation of 
the corresponding term in the free energy
with respect to $\pi D^2 L$ at fixed rod length $L$.) 
The latter result is known as Manning limiting law for osmotic coefficient of 
dilute solutions of weakly charged polyelectrolytes \cite{MAN69}.
This limiting law follows directly
 from Eq. (\ref{eq:freeapp1}) as we explain now.

For rod-like polyelectrolytes with $\xi\leq \xi_c$, the Manning limiting law 
states that the
osmotic coefficient, $\nu$, tends to a finite value 
as the solution becomes highly diluted,
\begin{equation}
  \lim_{D/R\rightarrow \infty} 
                             \nu=
                                (1-\frac{\xi}{2}).
\label{eq:mll}
\end{equation}
The osmotic coefficient here is defined as the ratio of the 
lateral osmotic pressure acting on the cell boundary, 
\begin{equation}
   \pi_{os}^{Lat}(D)=
                  -\frac{\partial{\mathcal F}_{elec}}
                        {2\pi D L\partial D}
\label{eq:latpi}
\end{equation} 
to the lateral osmotic pressure of an ideal gas of particles,
 $P_{id}$, filling the cell under similar conditions, {\em i.e.}
\begin{equation}
   \nu =
        \frac{\pi_{os}^{Lat}}{P_{id}}.
\label{eq:nu}
\end{equation}
Now, given the ideal-gas pressure 
$P_{id}=N_c/\pi L (D^2-R^2)$ and the (PB) electrostatic 
free energy of Eq. (\ref{eq:freeapp1}), 
one can easily recover the limiting law, Eq. (\ref{eq:mll}),
using Eqs. (\ref{eq:latpi}) and (\ref{eq:nu}).

\subsubsection{Case with $\xi\geq \xi_c$.}

In this case, it follows from Eq. (\ref{eq:beta2}) that $\beta$
tends to zero (as $(\ln D/R)^{-1}$) when $D/R\rightarrow \infty$.
Using this, one can observe that the potential field of 
Eq. (\ref{eq:Solution2}) reduces to
\begin{equation}
        \psi(r)\approx 
                     2\ln \frac{r}{R} 
                     + 2\ln [1+(\xi-1)\ln \frac{r}{R}],
\end{equation}
and the PB free energy, Eq. (\ref{eq:betafree1}), to
\begin{equation}
       \frac{{\mathcal F}^{PB}}{N}
                                \approx 
                                -\frac{f}{\xi} \ln \frac{D}{R},
\label{eq:freeapp1b}
\end{equation}
as estimated heuristically in Eq. (\ref{eq:freeapp_above}) in the text.  
Clearly in this case when $\xi>\xi_c$ , unlike the 
case with $\xi<\xi_c$, electrostatic repulsions
and entropic effects have contributions of the same order of magnitude
in the total longitudinal pressure which acts on the rod. This 
result may be understood in terms of the counterion condensation
model \cite{MAN69} as explained in Section \ref{sub:Osm}. 

The Manning limiting law for the (lateral) osmotic coefficient
is now obtained (from Eqs. (\ref{eq:latpi}), (\ref{eq:nu}) and 
(\ref{eq:freeapp1b})) as
\begin{equation}
  \lim_{D/R\rightarrow \infty} 
                             \nu=
                                \frac{1}{2\xi},
\label{eq:mll2}
\end{equation} 
where $\xi \geq \xi_c$.

In our model where $R$ is not fixed but depends on $\xi$
(see Eq. (\ref{eq:R})), the preceding discussions hold 
when $D\gg r_c+R_0$.


\subsection{Large grafting densities ($D/R\approx 1$):}

In a cell model with fixed $R$, the threshold
Manning parameter $\xi_c$ tends to zero as $D/R\rightarrow 1$
 \cite{FUOSS}. 
Therefore, for finite values of the
Manning parameter $\xi$, we have to use Eqs. (\ref{eq:beta2})
and (\ref{eq:betafree1}) for $\xi\geq \xi_c$.

Starting from Eq. (\ref{eq:beta2}), we can find 
an approximate expression for $\beta$ in the limit of 
$D/R\rightarrow 1$. Defining $\delta=D/R-1$ and 
expanding Eq. (\ref{eq:beta2}) for small $\delta$, we obtain
\begin{equation}
    \beta^2\approx 
               \xi (\frac{1}{\delta}+
                        {\mathcal O}(\delta^0))-1.
\end{equation}
Now replacing $\beta^2$ in Eq. (\ref{eq:betafree1}) and 
expanding in terms of $\delta$, we have (up to an
additive constant independent of $\delta$)
\begin{equation}
    \frac{{\mathcal F}^{PB}}{N}\approx  
                                -f\ln \delta +
                                 {\mathcal O}(\delta).
\label{eq:freeapp2}
\end{equation}
This is the entropic free energy of an ideal gas of particles
(Eq. (\ref{eq:identropy})) up to the leading order, and shows that
in this limit, the main contribution to the 
PB free energy comes from the entropy of 
counterions.  

In our model with the constant volume constraint, this limit 
(namely $D\rightarrow r_c+R_0$) has to be 
handled with care. In fact, the upper limit on $\xi$, that is $\xi_u$,
tends to the lower limit $\xi_0$, and so does the optimal Manning 
parameter. The threshold $\xi_c$ 
becomes smaller than $\xi_0$, 
therefore, the system indeed satisfies $\xi>\xi_c$ condition 
\cite{Note1}, so that the
above discussion remains valid.




\bibliographystyle{}

\begin{thebibliography}{10}



\bibitem{MIK88} S.J. Miklavic, S. Marcelja, J. Phys. Chem. 
{\bf 92}, 6718 (1988).

\bibitem{MIS89} S. Misra, S. Varanasi, P.P. Varanasi, Macromolecules
       {\bf 22}, 4173 (1989). 

\bibitem{PIN91} P. Pincus, Macromolecules {\bf 24}, 2912 (1991).

\bibitem{BOR91} O.V. Borisov, T.M. Birstein, E.B. Zhulina, 
   J. Phys. II (Paris) {\bf 1}, 521 (1991).

\bibitem{ROS92} R.S. Ross, P. Pincus, Macromolecules {\bf 25}, 2177
  (1992).

\bibitem{ZHU92} E.B. Zhulina, T.M. Birstein O.V. Borisov, J. Phys. II
  (Paris) {\bf 2}, 63 (1992).

\bibitem{WIT93} J. Wittmer, J.-F. Joanny, Macromolecules
{\bf 26}, 2691 (1993).

\bibitem{ISR94} R. Isra\"els, F.A.M. Leermakers, G.J. Fleer, E.B. Zhulina,
      Macromolecules {\bf 27}, 3249 (1994). 

\bibitem{BOR94} O.V. Borisov, E.B. Zhulina, T.M. Birstein, 
   Macromolecules {\bf 27}, 4795 (1994). 

\bibitem{PRY96} V.A. Pryamitsyn, F.A.M. Leermakers, G.J. Fleer,
   E.B. Zhulina, Macromolecules  {\bf 29}, 8260 (1996). 

\bibitem{ZHU97} E.B. Zhulina, O.V. Borisov, J. Chem. Phys. 
   {\bf 107}, 5952 (1997).

\bibitem{AMOS95} V.M. Amoskov, V.A. Pryamitsyn, Polymer Science 
USSR {\bf 37}, 1198 (1995). 

\bibitem{CSA01} F.S. Csajka, R.R. Netz, C. Seidel, J.-F. Joanny,
Eur. Phys. J. E {\bf 4}, 505 (2001).

\bibitem{CSA99} F.S. Csajka, C.C. van der Linden, C. Seidel, Macromol.
  Symp. {\bf 146}, 243 (1999). 

\bibitem{CSA00} F.S. Csajka, C. Seidel, Macromolecules {\bf 33}, 2728
  (2000);  Macromolecules {\bf 38}, 2022 (2005).

\bibitem{CS02} C. Seidel, Macromolecules {\bf 36}, 2536 (2003); 
Macromolecules {\bf 38}, 2540 (2005).

\bibitem{MIR95} Y. Mir, P. Auroy, L. Auvray, Phys. Rev. Lett.
  {\bf 75}, 2863 (1995).

\bibitem{GUE95} P. Guenoun, A. Schlachli, D. Sentenac, J.M. Mays,
  J.J. Benattar, Phys. Rev. Lett. {\bf 74}, 3628 (1995).

\bibitem{AHR97} H. Ahrens, S. F\"orster, C.A. Helm, Macromolecules
  {\bf 30}, 8447 (1997). 

\bibitem{AHR98} H. Ahrens, S. F\"orster, C.A. Helm,
  Phys. Rev. Lett. {\bf 81}, 4172 (1998). 

\bibitem{MUL00} F. Muller, M. Delsanti, L. Auvray, J. Yang, Y.J. Chen,
 J.W. Mays, B. Dem\'e, M. Tirrell, P. Guenoun, Eur. Phys. J. E {\bf 3},
  45  (2000).

\bibitem{MUL01} F. Muller, P. Fontaine, M. Delsanti, L. Belloni, J. Yang, 
 Y.J. Chen,  J.W. Mays, P. Lesieur, M. Tirrell, P. Guenoun, Eur. Phys. J. E 
 {\bf 6}, 109 (2001).

\bibitem{GUE98}   P. Guenoun, F. Muller, M. Delsanti, L. Auvray,
 Y.J. Chen,  J.W. Mays, M. Tirrell, Phys. Rev. Lett. {\bf 81}, 3872 (1998).

\bibitem{BAL02} M. Balastre, F. Li, P. Schorr, J. Yang, 
J.W. Mays, M.V. Tirrell, Macromolecules {\bf 35}, 9480 (2002).  

\bibitem{ABE02} S. Hayashi, T. Abe, N. Higashi, M. Niwa, K. Kurihara, 
Langmuir {\bf 18}, 3932 (2002).

\bibitem{HELM}  H.  Ahrens, S. F\"orster, C.A. Helm, N.A. Kumar, A. Naji,
R.R. Netz, C. Seidel, J. Phys. Chem. B {\bf 108}, 16870 (2004). 

\bibitem{GUENOUN}  G. Romet-Lemonne, J. Daillant, P. Guenoun, J. Yang, J.W. Mays, 
Phys. Rev. Lett. {\bf 93}, 148301 (2004). 

\bibitem{NAP83} D.H. Napper, {\it Polymeric Stabilization of 
   Colloidal Dispersions} (Academic Press, New York, 1983).

\bibitem{PAR97} Y.S. Park, Y. Ito, Y. Imanishi, 
    Chem. Mater. {\bf 9}, 2755 (1997).

\bibitem{ALFREY} T. Alfrey, P.W. Berg, H. Morawetz, J. Polymer Sci. 
{\bf 7}, 543 (1951).

\bibitem{FUOSS} R.M. Fuoss, A. Katchalsky, S. Lifson, 
Proc. Natl. Acad. Sci. USA {\bf 37}, 579 (1951).

\bibitem{KATCH} A. Katchalsky, Pure Appl. Chem. {\bf 26}, 327 (1971).

\bibitem{HOL01} C. Holm, P. K\'ekicheff, R. Podgornik (Editors), 
{\em Electrostatic Effects in Soft Matter and Biophysics} 
(Kluwer Academic Publishers, Dordrecht, 2001).

\bibitem{Naji03} A. Naji, R.R. Netz, C. Seidel, Eur. Phys. J. E {\bf 12}, 223 (2003). 

\bibitem{LIF} S. Lifson, A. Katchalsky, J. Polymer Sci. {\bf 13}, 43 (1954).

\bibitem{NET99b} R.R. Netz, H. Orland, Eur. Phys. J. E {\bf 1}, 67
 (2000). 

\bibitem{Note3} The electroneutrality condition entails that 
the canonical free energy, Eq. (\ref{eq:PBfree}), be invariant 
under the gauge transformation 
$\psi\rightarrow \psi+\psi_0$. As a consequence, the PB free energy (see Eqs.
(\ref{eq:betafree}) and (\ref{eq:betafree1})) 
is independent of $\kappa^2$.  

\bibitem{Note1} It is easy to check that 
the threshold Manning parameter, 
$\xi_c$, tends to the so-called Manning critical value $\xi_M=1$ (the onset 
of counterion condensation \cite{OOS71,MAN69}), when $D\rightarrow \infty$,
or $R_0$ and $r_c\rightarrow 0$.
In our model, the threshold Manning parameter 
$\xi_c$ may be smaller or larger than 
$\xi_0$, but it never exceeds $\xi_u$ 
defined in Eq. (\ref{eq:upxi}), {\em i.e.} $\xi_c<\xi_u$. 
Also it never becomes larger than one as
it follows easily from Eq. (\ref{eq:thresh}). Therefore, in a system with
$\xi_0>1$, we will always have the above-threshold condition 
$\xi\geq \xi_0>\xi_c$
implied by Eq. (\ref{eq:limitonxi}). Similar situation occurs
when $D\rightarrow R_0+r_c$, since in this case we have 
$\xi_u\rightarrow \xi_0$, and hence, again $\xi_c$ lies below 
$\xi_0$.

\bibitem{OOS71} F. Oosawa,  {\it Polyelectrolytes} (Marcel Dekker, 
   New York 1971).

\bibitem{MAN69} G.S. Manning, J. Chem. Phys. {\bf 51}, 924 (1969). 

\bibitem{ZIMM} M. Le Bret, B.H. Zimm, Biopolymers {\bf 23}, 287 (1984). 

\bibitem{NET98} R.R. Netz, J.-F. Joanny, Macromolecules {\bf 31}, 5123
(1998).

\bibitem{Note_elas} Replacing the FJC elasticity with the Gaussian 
chain elasticity given in Eq. (\ref{eq:gausselas}),  
the brush height is found to increase logarithmically 
by decreasing the grafting density in the low-grafting-density regime. 
For instance, we obtain $L_*/L_0\sim -\ln \rho_{\mathrm{a}} b_0^2$
for highly charged chains. Note that the elasticity model affects 
the magnitude of the stretching of the chains and also 
the particular dependence of the brush height on the grafting density
(although it can not produce a grafting-density dependence by itself \cite{AMOS95}). 
For chains of finite length, the correct limiting dependence
is obtained if we take into account the finite extensibility of the chains,
which is mimicked by the FJC model. 

\bibitem{SPE} R. Strebel, R. Sperb, Mol. Simul. {\bf 27}, 61 (2001).

\bibitem{ARNOLD} A. Arnold, C. Holm, Comput. Phys. Commun. {\bf 148}, 327 (2002). 

\bibitem{VMD} W. Humphrey, A. Dalke, K. Schulten, J. Molec. Graphics 
{\bf 14}, 33 (1996).

\bibitem{CS_unpub} C. Seidel, unpublished. 

\bibitem{NET01} R.R. Netz, Eur. J. Phys. E {\bf 5}, 557 (2001).


\end{thebibliography}

\end{document}